# Virtual Theranostic Trials: New Approach Methodologies (NAMs) and Precision Radiopharmaceutical Therapies

Carlos F. Uribe[1,2,3], Hamid Abdollahi[2,3], Tahir Yusufaly[4], Taylor J. McColl[2,3], Maziar Sabouri[2,5], James Fowler[2,5], Fereshteh Yousefirizi[2,3], Babak Saboury[2,6], Pedro L. Esquinas[1,3], Arman Rahmim[1,2,3,5]✉

1. Department of Molecular Imaging and Therapy, BC Cancer, Vancouver, BC, Canada
2. Department of Basic and Translational Research, BC Cancer Research Institute, Vancouver, BC, Canada
3. Department of Radiology, University of British Columbia, Vancouver, BC, Canada
4. Department of Radiology and Radiological Sciences, Johns Hopkins, Baltimore, MD, USA
5. Department of Physics & Astronomy, University of British Columbia, Vancouver, BC, Canada
6. Institute of Nuclear Medicine, Bethesda, MD, USA

✉ Correspondence:

Arman Rahmim, PhD, DABSNM, FSNMMI
675 West 10th Ave.; Office 6-112
Vancouver, BC V5Z 1L3
Canada
arman.rahmim@ubc.ca

## Abstract

Radiopharmaceutical therapies are expanding rapidly, but clinical evidence generation is limited by operational constraints and biological heterogeneity in radiopharmaceutical delivery and radiation risk. Virtual theranostic trials can run as companion trials, linking quantitative imaging to patient-specific models to generate evidence for personalized injections and scheduling, accelerate development, and broaden access.

**Introduction**

Theranostics, the pairing of quantitative molecular imaging with radiopharmaceutical therapies (RPTs), is expanding at a blistering pace, driven by new radiopharmaceutical classes, broader indications, and major biopharma investments. Market analyses forecast a multiple-fold increase in the sector in this decade (>USD 20.6 billion by 2030)[1].

RPTs also expose a deeper challenge; imaging modalities enable personalization faster than standard clinical trials can make it actionable. Whole-body quantitative imaging reveals inter- and intra-patient heterogeneities in radiopharmaceutical delivery, retention, and clearance, alongside dynamic tumor-organ interactions. However, conventional development pathways are optimized for population-average evidence, not for repeated, decision-relevant adaptations within individual patients.

This mismatch is now unavoidable in practice. Trials increasingly must answer questions that are central to precision RPTs. How should administered radioactivity be individualized? When should cycles be delayed, intensified, or discontinued? Who can safely tolerate absorbed dose escalation, and who requires de-escalation? Which combinations are rational, and when?

Model-informed drug development (MIDD), together with the broader emergence of new approach methodologies (NAMs) for faster and more efficient approvals, are reshaping how evidence is generated and interpreted, supported by formal regulatory pathways for early engagement on modelling strategies and context of use[2,3]. However, RPTs create a distinct opportunity because imaging is not just a biomarker, it is a quantitative readout of delivery and exposure that can anchor mechanistic prediction and uncertainty. This sets the stage for Virtual Theranostic Trials (VTTs), trial-embedded virtual counterparts that translate quantitative imaging into credible evidence with quantified uncertainty for adaptive RPTs.

**Virtual Theranostic Trials**

VTTs are pre-specified *in silico* trials designed specifically for theranostics that run in parallel with real clinical trials. Unlike retrospective analyses or generic quantitative systems-pharmacology simulations, VTTs are designed to guide trial decisions. They transform imaging, laboratory data, and clinical covariates into patient-specific mechanistic predictions, then aggregate these across virtual cohorts to anticipate population heterogeneity, rare subgroups, and variable care settings. Therefore, VTTs can evaluate practical questions, such as imaging schedules, theranostic digital twinning techniques, and predictive modeling strategies, under a single, decision-oriented framework.

A critical and often underappreciated prerequisite for VTTs to support longitudinal dose-response inference or adaptive between-cycle decisions, is lesion-consistent imaging across cycles. These models assume stable correspondence of tumors and organs over time, yet real-world imaging workflows often violate this assumption. Without reliable lesion and organ matching, key inferences become structurally ill-posed. In this context, artificial intelligence- (AI) enabled segmentation and lesion tracking, particularly for extensive disease, are not only downstream conveniences; they are upstream requirements for credibility.

The VTT platform and evidence loop comprise two layers (Figure 1).

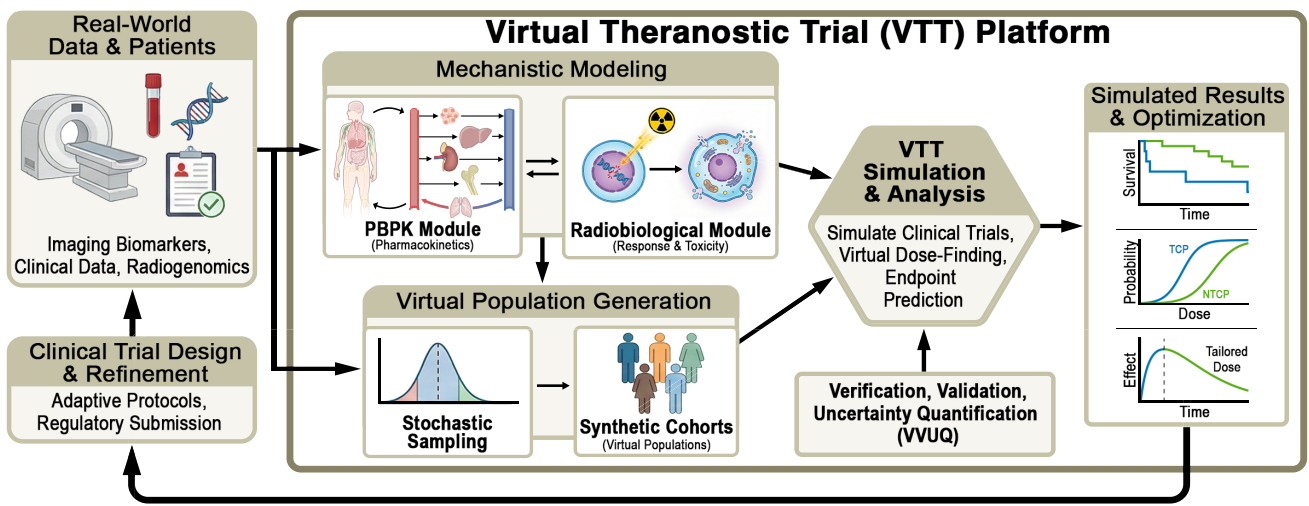

**Figure 1:** The VTT platform. Real-world imaging and clinical data parameterize the virtual population. Verification, validation, and uncertainty quantification ensure the VTTs are credible, supporting trial design and adaptive radioactivity injections and scheduling decisions.

First, mechanistic modeling built out of two submodules: Physiologically Based radioPharmacoKinetic (PBPK) modeling and radiobiological effects modeling. The first submodule, PBPK, links patient anatomy and physiology to radiopharmaceutical transport, binding, and clearance. PBPK is not simply curve-fitting; it makes explicit why two patients with similar tumors can experience radically different organ exposure, and it provides interpretable results for adaptation (e.g., receptor density, perfusion, tumor burden, renal function)[4]. The second submodule, radiobiology, allows dosimetry to become decision making by connecting radiation delivery to biological effect and toxicity, with uncertainty propagated end-to-end. The field of computational nuclear oncology (CNO) which we have promoted encompasses both dimensions[5]. The goal is not to replace clinical endpoints, but to generate prospective, testable predictions about dose-response, dose-toxicity, and schedule dependence. VTTs are most valuable when they are built around decisions clinicians face between cycles: continue, adapt, combine, or stop, bounding plausible outcomes under alternative actions at clinically relevant decision points.

Second, a virtual population. VTTs must represent heterogeneity intentionally, not only tumor phenotypes and organ reserve, but also real-world variation in imaging schedules, reconstruction, and care pathways. Synthetic cohort generation constrained by real-world data can enable testing of administration regimes and AI tools in populations that real trials under-sample (e.g. rare phenotypes, remote settings, under-represented groups). VTTs can become not only faster, but also more inclusive by design.

Implemented rigorously, VTTs shift evidence generation from an "average patient" paradigm toward distributions of outcomes conditioned on measurable patient factors and quantified uncertainty. This is a framing that aligns naturally with MIDD and emerging NAMs for evidence generation in oncology.

**Credibility before convenience**

VTTs will only matter if they are trusted. For simulations to inform clinical development or between-cycle decisions, credibility evidence must be risk-based and aligned to the context of use; what decision is being supported, how heavily the model is relied upon, and the consequences of being wrong. Verification, validation, and uncertainty quantification (VVUQ) should therefore be treated as first-class trial deliverables, not technical appendices[6–8].

Theranostics has a strategic advantage: models can be confronted against quantitative imaging, not only downstream outcomes. But this advantage is lost if VTTs are treated as a speed strategy. Since imaging protocols, reconstruction, and follow-up timing vary across centers, VTT credibility depends not only on model fidelity but also on robustness to real-world imaging variability. In practice, this means defining uncertainty budgets that explicitly include acquisition and reconstruction variability, missing time points, and longitudinal alignment errors, and propagating these uncertainties to decision-relevant quantities.

AI can lower barriers to VTT deployment, but it cannot substitute for credibility. Neural surrogates that accelerate modeling are valuable especially if they preserve traceability to mechanistic drivers and carry uncertainty into clinically meaningful tolerances. Similarly, AI for segmentation, lesion tracking, and quality control should be evaluated as part of the VTT evidence chain, because failures upstream can invalidate inferences downstream. The goal is not speed at any cost, but credibility-scored speed; faster computations with transparent assumptions, quantified uncertainty, and pre-specified decision limits.

A practical precedent comes from the FDA's VICTRE virtual imaging trial, which showed that *in silico* pipelines can reproduce clinically meaningful conclusions when validation datasets, model limits, and uncertainty are explicitly characterized[9,10]. Theranostics now needs an analogous ecosystem with shared reference datasets, benchmark tasks, and credibility targets tailored to common contexts of use such as absorbed dose verification, activity prescription, schedule optimization, and rare-subgroup extrapolation. Framed this way, VTTs become a theranostics-native NAM, not a replacement for trials, but a disciplined, audit-ready way to generate evidence that is faster to iterate and safer to generalize.

**A Short-Term Agenda**

To make VTTs operational, the theranostics community needs a concrete testable agenda:

1) Define the minimum viable VTT inputs. Specify the smallest imaging and clinical dataset that supports a patient model, and predefine acceptable uncertainty for each context of use (e.g., activity prescription, schedule adaptation, toxicity risk stratification).
2) Build auditable reference implementations. Containerized workflows, traceability of inputs and outputs, and open input schemas should be default. Open inspectable engines can accelerate convergence by making assumptions, parameterizations, and uncertainty propagation transparent. To this end, we are releasing the Python-based Computational Nuclear Oncology (PyCNO) library ([GitHub.com/Qurit/PyCNO](GitHub.com/Qurit/PyCNO)) as a dedicated computational engine for VTTs.

3) Enable prospective validation in real trials. VTTs should not be retrospective. They should generate pre-registered predictions that are confronted with subsequent cycles, toxicity, and response, with explicit criteria for model updating versus model failure.
4) Engage regulators early. Use established MIDD engagement pathways to align on context of use, credibility evidence, and decision limits before simulation results are used as supported evidence[3].
5) Treat access as a design constraint. VTTs should explicitly model heterogeneous care settings, not only ideal protocols, so recommendations remain safe and useful when imaging resources, timing, or tracer supply differ.

With risk-based VVUQ, VTTs can transform adaptive RPT evidence-generation into a scalable, auditable, and inclusive enterprise.

**Competing Interests**

B. Saboury is co-founder of United Theranostics. C. Uribe and A. Rahmim are co-founders of Ascinta Technologies, and P. Esquinas is employee of Ascinta Technologies.

**Funding:**

This work was in part supported by the Canadian Institutes of Health Research (CIHR) Project Grants PJT-180251 and PJT-197861.

**Author Contributions**

HA coined the term VTT, and along with TY, elaborated the multiple layers of VTTs. BS, CU, TY and AR developed ideas behind computational nuclear oncology and theranostic digital twins, and along with TM, MS and JW and PE contributed to creation of VTT modules. FY contributed to the role of AI for VTTs. CU and AR supervised overall creation of VTT workflows. All authors contributed significantly to the writing of the manuscript.